\documentclass[12pt]{iopart}
\usepackage{epsfig}
\usepackage{graphicx}% Include figure files
\usepackage{dcolumn}% Align table columns on decimal point
\usepackage{bm}% bold math
\begin{document}
\title{\Large A study of gravitational collapse with decaying of the vacuum energy}
\author{M. de Campos}
\address{(1) Physics Department, Roraima Federal University,
Campus do Paricar\~{a}na, Bloco III, Boa Vista, RR, Brazil}
\ead{campos@on.br}
\begin{abstract}
We study  the gravitational collapse  of a dust dark matter, in a $\Lambda $-background.
We consider two distinct cases:  First we do not have a  dark matter and 
 dark energy coupling; second, we consider that $\Lambda $ decay in dark particles.
The approach adopted assumes a modified  matter expansion rate and we have
formation  of a  black hole, since that,  we have  the formation  of an
apparent horizon.  Finally,  a brief comparison of the  process of the
matter condensation using the  gravitational collapse approach and the
linear scalar perturbation theory is considered.
\end{abstract}
\pacs{04.70.Bw}
\maketitle
\section{Introduction}
In the course of the eighteen century, astronomers were discussing the internal structure of the
individual nebulae rather than the wider cosmological problem.  Probably, one of the obstacle to
study the wider context is related to the question: Is the universe finite?

On the other hand, the infinitely extended universe considered by the Newton's theory of
gravitation was one of the reasons for the believe that Newton's theory was cosmologically
unacceptable.

The advent of the special and general theory of relativity by A. Einstein furnish a new basis to
discuss the wider cosmological problem. We mention two important discoveries about the universe
that emerge from the general relativity.  First, follow  Poisson  and  Israel  \cite{Israel},
``black hole  theory  is without  any  doubt  one  the  major  triumphs  of  classical  general
relativity''. Although, based in  the existence of a spacelike  singularity that destroys quantum
information  and in the  presence   of  an  infinite  red-shift  surface, it was speculate that black
holes may not exit at all \cite{Chapline}.  Second,  one of  the recent  important discoveries about the
cosmos  is the accelerating  expansion  of  the cosmic fluid. 
In order to obtain a universe with an  accelerated expansion we must provide a new
component for the energy distribution, that is usually referred as dark energy.
One  of  the  candidates responsible for this process  is the cosmological constant, frequently
denoted by $\Lambda $.

Only few  years after Albert  Einstein introduced the  fundamentals of
the general relativity theory, himself includes in the field equations
of the general relativity theory the cosmological constant.  Moreover,
in the light of the  experimental evidence of the universal expansion,
Einstein  considered  the  inclusion  of  $\Lambda$  ``...the  biggest
blunder of my life.''\cite{Gamov}.

Before  the   results  from  supernova of type IA observations   appear  in  the
literature, that  indicates an accelerated expansion  of the universe,
L. Krauss  and M.  Turner  call our attention that  ``The Cosmological
Constant  is  Back  ''.  They  cited  the  age  of the  universe,  the
formation of the  large scale structure and the  matter content of the
universe as the data that  indicates the insertion of the cosmological
constant \cite{Krauss}.
The  cosmic   microwave  background  radiation
anisotropy and large scale structure, also indicates this acceleration
process of the universe  \cite{Riess}, \cite{Perlmutter}.  

The mechanism that triggered the  acceleration of the universe has not
been identified, and the simplest  explanation for the this process is
the inclusion  of a non null cosmological  constant.
However,  the  inclusion  of  the cosmological  constant  creates  new
problems. Some of them are  old, as the discrepancy among the observed
value  for the  energy  density of  the  vacuum, and  the large  value
suggested   by   the    particle   physics   models   \cite{Weinberg},
\cite{Garrig}.  In spite of of the problems caused by the inclusion of
$\Lambda $, the cosmological scenario with $\Lambda$  has a good agreement in respect
to  the age  of the  universe  estimate, the  anisotropy of  microwave
background radiation and the supernova experiments. Beyond this, making
several assumptions concerning with the spectrum of fluctuations in the 
early  universe and  the formation of galaxies,  G. Efstathiou suggests 
that the small value  of  the  cosmological constant  may be explained by 
the anthropic principle \cite{Efstathiou}.

The observational  data also suggest  that the universe is  flat.  The
flatness is consequence  of the cosmic content and  indicates that our
universe consists of $\frac{2}{3}$ of dark energy and $\frac{1}{3}$ of
dark  matter, approximately. The dark matter component dominates at small 
scales ($< 500 Mpc$) and the dark energy component dominates at large 
scale ($> 1000 Mpc$).

The evidence that these new components of the universe are different 
substances has been considered in literaure \cite{Sandvik}.  The component 
at small scale, generally, is considered as weakly interacting massive particles 
and the component at large scale is associated to some form of a scalar field.
A link between both components to a scalar field is study by Padmanabhan and Choudury 
\cite{Padmanabhan}.  Consequently both components have a scale-dependent state equation, 
but. today, both  components are  unknown  is respect  to the  your
nature. 

The presence  of the dark matter was  fist discussed by Zwiecky
\cite{Zwiecky} in  your study of  the dynamics of galaxy  clusters. At
the present  time the presence of  the dark matter is  indicate by the
study of  galaxy rotation curves,  the structure of the  galaxy groups
and clusters, large scale cosmic flows and gravitational lensing. This
last, a phenomenon proposed by Zwiecky himself as an astronomical tool
\cite{Zwiecky1}.

An alternative  model that  furnish a negative pressure in  the cosmic fluid and
results in an  accelerated expansion of the universe  is known as open
system cosmology \cite{Prigogine}.  In this model the particles number
of  the universe  do not  conserve and  the energy-momentum  tensor is
reinterpreted  in the  Einstein's equations.   This extra  pressure is
known as creation pressure and is negative \cite{Alcaniz},\cite{Lima}.
The creation process is due to the expenses of the gravitational field
and is an irreversible process.  One of the attractive features of the
hypothesis of  particle production is  the relation between  the large
scale   properties   of  the   universe   and   the  atomic   phenomena
\cite{McCrea}.

If the final state of a gravitational collapse results a black hole or
a naked  singularity, it is necessary  to look for  the development of
apparent horizons and to examine if there are any families of outgoing
non-spacelike  trajectories,  which  terminates  in the  past  at  the
singularity.   The apparent  horizon, that  is formed  within  the star,
characterise  a   trapped  surface  region  of   the  spacetime.   The
singularity evolves to a naked  singularity if the neighbourhood of the
star center  do not gets trapped  earlier than the  singularity.  In
this  case the non  space-like future  trajectories escaping  from the
center of the star  to outside observers \cite{Joshi}.  Otherwise, the
black hole is  formed if we have a region  around the neighbourhood 
the star center gets trapped earlier than the singularity.

The natural question,  how the dark energy effects  the process of the
gravitational  collapse was  studied by  several  authors, \cite{Cai},
\cite{Ferreira},\cite{Mota}, \cite{Lokas}.  Cai  and Wang consider the
collapse  of two  component fluid  constituted  by a  dust cloud  dark
matter coupled  with a  dark energy component.   They assume  that the
interaction between the two components is given by:
           \begin{equation}
       \frac{\rho _{\Lambda}}{\rho _{dm}}=A R^{3n}\, ,
     \end{equation}
where $A$ and  $n$ are constants.  They conclude  that, in presence of
dark energy  the gravitational collapse  of the dark matter  cloud can
occurs.  Although, the  junction of the star to  the spacetime outside
is not considered by the  authors and, in this  work, we assume
identical approach.

Resuming, in this  work our intention is  study the gravitational  collapse of a
cloud of dark matter in a background with a cosmological constant, where we have particle production
of dark matter particles  at the expense of the dark energy backgound .
\section{The collapse of a dark matter cloud}
We can divide the spacetime of the star into three regions. One correspondent to the
interior of the star, one to the exterior and one relative to a spherical surface, that divide the
interior  of the exterior. The motion of this surface is described  by a  timelike
three dimensional space $\Sigma $,  given by $r_\Sigma =  constant$. The metric  on $\Sigma $
can be  written as $ds^2  _{\Sigma}=d \tau ^2 -R(\tau)^2  d\Omega ^2$,
where  we  use  the   intrinsic  coordinates  $\xi  ^a  \equiv  (\tau,
\theta,\phi)$, $a(\tau)=r_{\Sigma}R(\tau)$ and $\tau =t$.
Once the spacetime inside the surface is fixed, whether a thin shell appears on $\Sigma$ or
not is completely determined by the spacetime outside the star \cite{Cai}.

Our  interest in  the interior  of the  star is  due to  a  coupled of
reasons.  First, as  discussed by Cai and Wang  \cite{Cai}, we do not have a unique matching
of exterior and the interior spacetimes.
Second, the discussion about the  formation of black holes, reduces to
the one whether apparent horizons develop inside the star.

We assume that the spacetime inside the star is homogeneous and isotropic.
Consequently, the interior of the star is described by
   \begin{equation}
      ds^2=dt^2-R(t)^2(dr^2+rd\Omega ^2) \, ,
       \end{equation}
where $d\Omega ^2 = d\theta ^2+\sin{\theta}^2d\phi ^2$.
 
The formation of a black hole creates an apparent horizon, hence
       \begin{equation}
               a_{,\alpha}a_{,\beta}g^{-\alpha \beta}= [r\dot R(t)]^2-1=0 \, ,
                  \end{equation}
where $a(r,t)=rR(t)$ is the geometrical radius of the two-sphere $t ,r
=constant$ and 
the dot means the time derivative.

Cahill and Mac Vittie \cite{Cahill} using curvature coordinates and considering the metric tensor
continuous across the surface of discontinuity, which separates a spherical distribution of material
from the surrounded empty space, find a function that may be defined as the total mass-energy
entrapped inside the radius $r$ at the time $t$, namely
              \begin{equation}
                  m(t,r)=\frac{1}{2}a(1+a_{,\alpha}a_{, \beta}g^{\alpha \beta})=\frac{1}{2}r^3R\dot R ^2 \, .
                      \end{equation}
If we consider that the all star collapses inside the apparent horizon, using (3) we have
   \begin{equation}
        \dot a ^2 (\tau _c)\mid _{\Sigma} = 1 \, ,
             \end{equation}
where $t_c$ is the time necessary to occur the collapse.  

Therefore, on the surface of the star, eq. (4) furnish the total mass of the collapsed
star, namely
     \begin{equation}
             M=\frac{1}{2}a(\tau ){\dot{a}(\tau)}^2\, ,
                \end{equation}
and the mass of the black hole formed is $M_{BH}=M(\tau _c)$.

  The energy momentum tensor inside the star is given by
    \begin{equation}
            T_{\mu \nu}=(\rho_\Lambda +\rho_{dm} +P_\Lambda)u_\mu u_\nu \,,
                 \end{equation}
where $\rho_\Lambda$ and $P_\Lambda$ are the density and the pressure of dark energy and
$\rho_{dm}$ is the dark matter density, while $u_\mu$ is their four velocities.

Taking into account that the reference system  is just the matter filling  the
space, the Einstein field equations are:
       \begin{eqnarray}
              \frac{\ddot R}{R}&=&-\frac{1}{6}\kappa (\rho_\Lambda +\rho_{dm}+3P_{\Lambda})\, , \\
                   \frac{\dot{R}^2}{R^2}&=&\frac{1}{3}\kappa(\rho_\Lambda +\rho_{dm})\, .
                       \end{eqnarray}
The interaction between the dust matter and dark energy follows from the conservation law
   \begin{equation}
       T_{\mu \nu ;\lambda}g^{\nu \lambda}\, ,
          \end{equation}
which in the present study, taking into account the decay of $\Lambda$
into dark particles, assumes the form
     \begin{equation}
          \dot{\rho}_{dm}+3\frac{\dot R}{R}\rho_{dm}=-\dot{\rho}_\Lambda\, .
                  \end{equation}
In the absence of the coupling  between dark matter and dark energy we
have $ \dot{\rho}_{dm}+3\frac{\dot R}{R}\rho_{dm}=0$, which integration results  $\rho _{dm}=\rho_{dm0}R^{-3}$, 
 where $\rho_{dm0}$ is an integration constant.

Considering that  the dark  energy decay into  cold dark  matter, this
will  dilute in  a smaller  rate  compared with  their usual  relation
proportional to $R^{-3}$ \cite{Meng}.  Hence
                 \begin{equation}
               \rho_{dm}=\rho_{dm0}R^{\epsilon -3}\, ,
            \end{equation}
where $\epsilon $ is a  positive constant, that indicates the deviation
from the absence of the coupling between the dark energy and dark matter.

We can write equation (11) as
      \begin{equation}
          \frac{d\rho_{dm}}{dR}+3\frac{\rho_{dm}}{R}=-\frac{d\rho_\Lambda}{dR}\, .
             \end{equation}

Substituting (12) into (13), we find the expression for the dark energy density, namely
      \begin{equation}
         \rho_\Lambda=\rho_{\Lambda 0}- \epsilon\rho_{dm 0}\frac{R^{\epsilon -3 }}{3-\epsilon}\, .
           \end{equation}

Our intent is study the gravitational collapse, then the condition $\dot{R}<0$ must be considered.

Note that, in the collapse  process the scale factor diminishes, hence
the  density  of   the  dark  matter  grows  and   energy  density  of
$\Lambda$-component     diminishes.  In addition, for $R\rightarrow     0$,
we have $\rho_{\Lambda}\rightarrow -\infty$.

The validity interval for $\epsilon$ is $\epsilon\leq 1$ in conformity
with SNe IA observation.  Otherwise, we have an accelerated universe in a matter dominated era. On
the other hand, the SNe IA data indicates that we have an decelerated expansion before
redshift  $z\approx 0.46 \pm 0.13$ \cite{Riess1}.   The evidence for
cosmic deceleration that precede the accelerated expansion, on another
words,  a  strong  evidence  for  a cosmic  jerk  \cite  {Visser},  is
inconsistent with very rapid evolution of dark energy. There isn't any
anomalous cold  dark matter expansion  rate observable. So,  we expect
that $\epsilon <<1$.

With the auxilious of equations (9) , (12) and (14) we find the equation
\begin{equation}
\dot{R}^2-K_I R^2-K_ { II}R^{\epsilon -1}=0 \, ,
\end{equation}
where $K_I = \frac{8\Pi G}{3}\rho _{\Lambda 0}$ and
$K_{II}=\frac{8\Pi G }{3}(1-\frac{\epsilon}{3-\epsilon})\rho_{dm0}$.

Although   we   have   a   vacuum   decaying,   the   state   equation
$w=\frac{P_\Lambda}{\rho_\Lambda}$ is still constant.
\subsection{Gravitational collapse of a dust cloud without decay of $\Lambda$}
In this section we consider the collapse of a dust cloud in the presence of a
cosmological constant without creation of dark particles. 
Therefore, $\epsilon =0$ and  equation (15) can be written as
\begin{equation}
\dot{R}^2-Q_I R^2-Q_ { II}R^{-1} =0\, ,
\end{equation}
where $Q_I = \frac{8\Pi G}{3}\rho _{\Lambda 0}$ and $Q_{II}=\frac{8\Pi G \rho_{dm0}}{3}$.

The integration of last equation results:
\begin{equation}
 R(t)=R_0\{\sinh{\frac{3\sqrt{Q_I}(t_0-t)}{2}}\}^{\frac{2}{3}}\, ,
\end{equation}
where $t_0$ is an integration constant and $R_0=(\frac{Q_{II}}{Q_{I}})^{\frac{1}{3}}$.

Consequently, the expressions for the dark matter and dark energy
densities are, respectively:
\begin{eqnarray}
 \rho_{dm}&=&\rho_{dm0}\,R_0^{-3}\{\sinh{\frac{3\sqrt{Q_I}(t_0-t)}{2}}\}^{-2} \, ,
\\
\rho_{\Lambda}&=&\rho_{\Lambda_0}\, . 
\end{eqnarray}
The other relevant pertinent quantities are:
\begin{eqnarray}
\dot{a}(t)&=&-a_0\sqrt{Q_I}\{\sinh{\frac{3\sqrt{Q_I}(t_0-t)}{2}}\}^{-{\frac
{ 1 } { 3 } } } \cosh { \frac { 3\sqrt {Q_I } (t_0-t) } { 2 } } \, , \\
 M(\tau) &=&\frac{1}{2} a_0^{3}\,Q_{I}\{\cosh{\frac{3\sqrt{Q_I}(\tau_0-\tau)}{2}}\}^2\, ,
\end{eqnarray}
where $a_0=rR_0$.

In the limit $t\rightarrow t_0$, the equations (17)-(21) reduces to
\begin{eqnarray}
 R(t)&=&\frac{3}{2}R_0\sqrt{Q_I}(t_0-t)^{\frac{2}{3}} \\
 \rho_{dm}&=&\rho_{dm0}R_0^3(t_0-t)^{-2}  \\
 \rho_{\Lambda}&=&\rho_{\Lambda_0}  \\
 \dot{a}(t) &=& -a_0 \sqrt{Q_I}(t_0-t)^{-\frac{1}{3}} \\
M(t)&=&\frac{1}{2}a_0 ^3 Q_I \, ,
\end{eqnarray}
that are identical to Oppenheimer-Synyder solution, considered as the first study of the
gravitational collapse \cite{Oppenheimer}.

Comparing the formation of the apparent horizon in the presence of
$\Lambda$ with Oppenheimer-Synyder solution, note that for the
first case the   apparent horizon appear in a time anterior that the
Oppenheimer-Synyder case, vide Fig. 1.
\begin{figure}[!ht]
{\includegraphics[width=3cm]{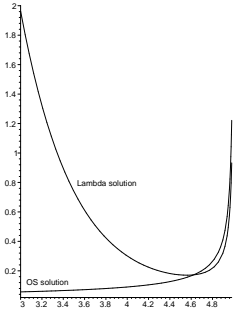}}
\caption{Evolution for ${\dot R}^2 $ considering the Oppenheimer-Synyder solution (OS) and the 
solution in the presence of $\Lambda $.}
\label{fig:Fig. 1}
\end{figure}

With respect to the total mass collapsed, we have a constant value for
the Oppenheimer-Synyder solution, given by the  Schwarzschild black hole mass. 

The spacetime singularity is formed at $\tau =  \tau_0$ and the apparent horizon is formed at
\begin{equation}
 \tau=\tau_0-(\frac{1}{Q_I a_0 ^2})^{-\frac{3}{2}}\, .
\end{equation}
Hence, the black hole is formed since that the apparent horizon is anterior to the reach of the
singularity and the star is not initially trapped. 
\subsection{Gravitational collapse of a dust cloud with decay of $\Lambda$}
Now, we consider the creation of dark particles at the expenses of the $\Lambda$ decay.
Consequently, the integration of eq.(15) results
              \begin{equation}
                            R(t)=R_0\{\sinh{\frac{\sqrt{K_I}(3-\epsilon)(t_0-t)}{2}}\}^{\frac{2}
                                  {3-\epsilon}} \, ,
 \end{equation}
where $t_0$ is an integration constant and $R_0=(\frac{K_{I}}{K_{II}})^{\frac{1}{\epsilon
-3}}$. Naturally this solution decay in (22) for $\epsilon =0$.

Considering (28) the physically important quantities are:
\begin{eqnarray}
&\rho _{dm}&=\rho _{dm0} R_0^{-3}\{\sinh{\frac{\sqrt{K_I}(3-\epsilon)(t_0-t)}{2}}\}^{-2}\, , \\
&\rho _{\Lambda} &=\rho _{\Lambda 0}+\frac{\epsilon}{\epsilon -3} \rho
_{dm0}[\sinh{\frac{\sqrt{K_I}(3-\epsilon)(t_0-t)}{2}}]^{-2} \\
&\dot{a}(t)&=-a_0\sqrt{K_I}[\sinh{\frac{\sqrt{K_I}(3-\epsilon)(t_0-t)}{2}}]^{\frac{\epsilon
-1}{3-\epsilon}}\\
&[&\cosh{\frac{\sqrt{K_I}(3-\epsilon)(t_0-t)}{2}}]\, , \nonumber \\
&M(\tau)&=\frac{K_I}{2}R_0^3[\sinh{\frac{\sqrt{K_I}(3-\epsilon)(\tau_0-\tau)}{2}}]^{\frac{ 2 \epsilon
}{3-\epsilon}}\\
&[&\cosh{\frac{\sqrt{K_I}(3-\epsilon)(\tau_0-\tau)}{2}}]^2 \, . \nonumber
\end{eqnarray}
The profile of dark energy density appear in the Figure 2.
\begin{figure}[!ht]
{\includegraphics[width=3cm]{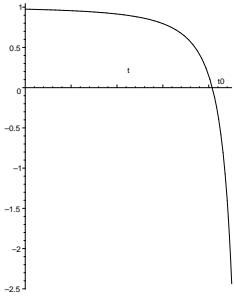}}
\caption{Evolution of the dark energy density for $\epsilon =0.1$.}
\label{fig:Fig. 1}
\end{figure}

It is interesting look to the change of $\Lambda $ signal. The
background change to an  anti-deSitter type spacetime in a time
anterior to the reach of the singularity that is given by:
\begin{equation}
 t=t_0-arcsinh{\{[\frac{\epsilon \rho_{dm0}}{(3-\epsilon)\rho_{\Lambda_0}}]^{\frac{1}{2}}\}} \, .
\end{equation}

Physically, a negative $\Lambda $ corresponds in the Newtonian limit to an extra attractive term in
the gravitational force.  The anti-de Sitter spacetime play an important role in the superstring
theory.  Probably, a scenario with superior dimensions is more adequate to study
this change of signal of $\Lambda$ and the consequences.

The evolution of the $\dot{R}^2$  is better analysed making a graph.
The profile for two different values for $\epsilon$ appear in the Fig.3.
\begin{figure}[!ht]
{\includegraphics[width=3cm]{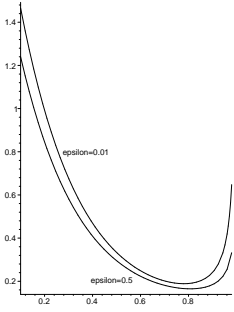}}
\caption{Evolution of the $\dot{R}^2$ in the presence of the $\Lambda$ decay into dark particles for
two different values for $\epsilon$.}
\label{fig:Fig. 1}
\end{figure}
Note that, with the increase of $\epsilon $, $\dot{R}^2$ reaches  the unity  
in an epoch anterior than smaller values for $\epsilon$. Using other
words, small values for $\epsilon$ favour the formation of the apparent
horizon.  In this case, identically to the anterior case, we have the formation of a black hole
 
In respect to the  mass inside the star we obtain an identical behaviour. Hence,
$M_{\epsilon=0.2}<M_{\epsilon=0.01}$.
Consequently, we hope that the increase of $\epsilon $ is harder for the process of matter
condensation.

A similar conclusion can be obtained studying the evolution of the scalar perturbations.
The equation that governs the evolution of the scalar linear perturbations in a matter dominated
phase is \cite{Lemos}
\begin{equation}
 R^2\frac{d^2\delta}{dR^2}+\frac{3}{2}(1+\frac{\Lambda}{3H^2})R\frac{d\delta}{dR}+(\frac{\Lambda}{
2H^2}-\frac{3}{2} )\delta=0 \, ,
\end{equation}
that assumes the form:
\begin{equation}
 R^2\frac{d^2\delta}{dR^2}+\frac{3+\epsilon}{2}R\frac{d\delta}{dR}+\frac{\epsilon -3}{2}\delta=0\, .
\end{equation}
The last equation was founded using eqs. (9), (12) and (14). In addition, is also necessary
assume that $\frac{\rho _{dm}}{\rho_{\Lambda}}=\frac{3-\epsilon}{\epsilon}$, that is equivalent to
$\rho _{\Lambda} << 0$.  This assumption with the auxilious of eqs. (9), (12) and (14), results
$\Lambda =\epsilon H^2$.

When the $\Lambda $ terms are ignored in eq. (34) we obtain the usual solution $\delta \propto R$

Integrating the eq. (35) we obtain
\begin{equation}
 \delta \propto R^{\frac{\sqrt{\epsilon ^2 -6\epsilon +25}-\epsilon -1}{4}}\, ,
\end{equation}
 which evolve slower than the usual solution for the density contrast proportional to $R$
\cite{Meng}.

\end{document}